\documentclass{article}

\usepackage{PRIMEarxiv}

\usepackage[utf8]{inputenc} 
\usepackage[T1]{fontenc}    
\usepackage{hyperref}       
\usepackage{url}            
\usepackage{booktabs}       
\usepackage{amsfonts}       
\usepackage{nicefrac}       
\usepackage{microtype}      
\usepackage{lipsum}
\usepackage{fancyhdr}       
\usepackage{graphicx}       
\graphicspath{{media/}}     
\usepackage{multirow}

\usepackage{array}

\title{Investigating Student Interaction Patterns with Large Language Model-Powered Course Assistants in Computer Science Courses}

\author{
  Chang Liu$^{1}$, Loc Hoang$^{2}$, Andrew Stolman$^{2}$, Rene F. Kizilcec$^{2,3}$, Bo Wu$^{1,2}$ \\
  $^{1}$Colorado School of Mines, Golden, USA \\
  $^{2}$HiTA AI Inc., Santa Clara, USA \\
  $^{3}$Cornell University, Ithaca, USA \\
  \texttt{liuchang@mines.edu}, \texttt{loc@hita.ai}, \texttt{astolman@hita.ai}, \texttt{kizilcec@cornell.edu},
  \texttt{bwu@mines.edu}\\
}

\begin{document}
\maketitle

\begin{abstract}
Providing students with flexible and timely academic support is a challenge at most colleges and universities, leaving many students without help outside scheduled hours. Large language models (LLMs) are promising for bridging this gap, but interactions between students and LLMs are rarely overseen by educators. We developed and studied an LLM-powered course assistant deployed across multiple computer science courses to characterize real-world use and understand pedagogical implications. By Spring 2024, our system had been deployed to approximately 2,000 students across six courses at three institutions. Analysis of the interaction data shows that usage remains strong in the evenings and nights and is higher in introductory courses, indicating that our system helps address temporal support gaps and novice learner needs. We sampled 200 conversations per course for manual annotation: most sampled responses were judged correct and helpful, with a small share unhelpful or erroneous; few responses included dedicated examples. We also examined an inquiry-based learning strategy: only around 11\% of sampled conversations contained LLM-generated follow-up questions, which were often ignored by students in advanced courses. A Bloom’s taxonomy analysis reveals that current LLM capabilities are limited in generating higher-order cognitive questions. These patterns suggest opportunities for pedagogically oriented LLM-based educational systems and greater educator involvement in configuring prompts, content, and policies.
\end{abstract}


\section{Introduction}
\label{sec1}

Traditionally, students engage in direct interactions with educators to acquire knowledge. These instructional interactions are grounded in educators' pedagogical training and experience, and are continuously monitored by them \cite{caires2012becoming}. However, such interactions are often insufficient to meet the diverse needs of individual students or to provide support outside of campus hours. Large language models (LLMs) have shown considerable capabilities across multiple domains including knowledge exploration and information retrieval \cite{chang2024survey, chen2024large, li2023evaluating}. They provide a promising bridge to address support gaps in traditional education. In current educational institutions, students are increasingly utilizing commercially available LLM products for information searching and instructional guidance \cite{intelligent2023gpt, chan2023students, grande2024student}. Despite being in their early stages, LLMs have already exerted a substantial influence \cite{extance2023chatgpt, wang2024large, grande2024student}.

Unlike traditional educator-student teaching patterns, interactions between students and LLMs typically rely on proactive engagement initiated by students. This shift reflects a different interaction pattern in which LLMs act as responsive agents that react to student questions and generate text-based answers. It introduces new and distinct challenges for students as they adapt to this emerging form of education. We therefore identify three key challenges: First, LLMs exhibit certain limitations, such as hallucination and reliance on potentially outdated training data \cite{huang2023survey, jia2024assessing, openai2024chatgpt, bubeck2023sparks}. Misinformation in the educational domain poses unique challenges as students may struggle to identify inaccuracies. Second, LLMs typically function as passive agents that respond to human questions; this new interaction pattern requires a higher level of user motivation and has not been thoroughly characterized \cite{amoozadeh2024student, lieb2024student}. Third, the pedagogical strategies for supervising human-LLM educational interactions remain understudied \cite{lee2024life}. This gap impedes the evolution of the current educational paradigm in the presence of LLMs.


Despite these challenges, LLMs can provide ubiquitous accessibility and real-time responses to user questions regardless of time and location. Building on these advantages, our primary contributions include:

\textbf{(1)} We have implemented an LLM-powered educational platform that serves as a course assistant. Our system is built on Retrieval-Augmented Generation (RAG) technology, incorporating multiple features that work collaboratively to provide a structured learning experience. The RAG component retrieves relevant context from an educator-curated database and delivers this information to the LLMs. Additionally, the customization feature enables the construction of course-specific knowledge databases and educator-defined response generation rules. Our multi-route question dispatching feature processes user queries across multiple modes, including general, homework, and practice modes. The system provides an open-ended, high-agency web interface that enables users to conduct learning activities according to their preferences.

\textbf{(2)} The system has been actively deployed since early 2023 across multiple schools and universities spanning different disciplines. As of Spring 2024, our system has been implemented across three higher education institutions and serves approximately 2,000 users. In this study, we specifically analyze interaction data from the Spring 2024 semester to study user-LLM interaction patterns. Our analysis reveals interaction patterns across multiple dimensions, including temporal usage and conversation mode-level analysis. Key findings indicate that (A) LLM-powered systems can effectively complement traditional education support gaps, (B) users heavily rely on LLMs for homework-related questions, suggesting the need for a specialized homework mode that provides hints rather than direct answers, and (C) strategically promoting user agency in educational scenarios may strengthen students’ learning initiative and foster higher engagement.

\textbf{(3)} When interacting with our system, users exercise considerable agency and take the initiative to complete the learning processes. We consider this process partially aligned with the well-established inquiry-based learning theory \cite{pedaste2015phases,abdi2014effect} and enhance our system by adding functions such as posing follow-up questions to users. Our results indicate that only a few LLM-generated questions are followed up by users, while others are typically ignored. This observation challenges the implementations of educational theories in LLM-based systems. It prompts further discussion on how to properly leverage LLMs to facilitate students' learning processes.

In summary, our study aims to answer the following research questions:
\begin{itemize}
\item RQ1: Does the course assistant address the gaps in traditional in-school education, either at the temporal or individual level?
\item RQ2: What is the distribution of scenarios in which students choose to use the course assistant? Is it primarily for general purposes, homework, or practice?
\item RQ3: What is the cognitive level of student-initiated questions, and, when applicable, of the LLM’s follow-up questions? Do the LLM’s follow-up questions help promote deeper cognition?
\item RQ4: What is the effect of applying inquiry-based teaching strategies to guide student–LLM interactions?
\end{itemize}

\section{Related Work}

\subsection{LLMs and Related Techniques}
LLMs are a type of large machine learning model that is pre-trained on vast amounts of data and can generate human-like language. They have demonstrated impressive performance in multiple areas, including human language generation, question answering, and information retrieval \cite{chang2024survey, chen2024large, li2023evaluating}. In recent years, LLMs and their applications have undergone rapid development \cite{dam2024complete, tan2024chat}. Among these, Generative Pre-trained Transformer (GPT) models are some of the most prominent LLMs, whose remarkable capabilities have led to several mature products, such as ChatGPT and Claude \cite{openai2024chatgpt, bubeck2023sparks, anthropic2024claude}.

Retrieval-Augmented Generation (RAG) is an efficient LLM-based technique that has proven effective in current LLM research \cite{li2022survey, lewis2020retrieval}. Systems based on RAG typically include a knowledge database that can be searched based on user questions using specific similarity metrics. The relevant knowledge context is then retrieved and provided to the LLMs, enabling them to generate content aligned with the provided context. Utilizing RAG allows LLMs to access the latest or domain-specific data that were not available during pre-training or fine-tuning. Given these advantages, RAG has been widely adopted in a variety of LLM applications, including those in education \cite{siriwardhana2023improving, gu2018search, liu2024teaching}.

\subsection{LLMs in Education: Applications and Concerns}
Particularly, research has been conducted to explore the integration of LLMs into educational contexts \cite{extance2023chatgpt, jeon2023large}. Students in higher education have actively used LLMs for learning assistance \cite{chan2023students}, although concerns have been raised about the unprecedented influence these interactions may have on learners \cite{kasneci2023chatgpt, zhai2022chatgpt, amoozadeh2024student}. More critically, these interactions often lack consistent supervision by educators. This situation raises questions about the quality and appropriateness of the assistance provided by LLMs \cite{cardona2023artificial}.

Prior research has demonstrated the efficacy of LLM-powered systems in functioning as course assistants. For instance, Pardos et al. introduced an LLM-based tutoring system capable of generating tailored scaffolds and hints for students in algebra courses \cite{pardos2023oatutor}. Similarly, Kazemitabaar et al. developed CodeAid, an LLM-driven tool that provides feedback and suggestions to students learning computer programming \cite{kazemitabaar2024codeaid}. Their research included comprehensive interviews and surveys involving both educators and students. This approach yielded detailed insights into the system's effectiveness. Another notable system, Jill Watson, is an LLM-based AI that serves as a course assistant, responding to students' questions about course content in online discussion forums \cite{eicher2018jill, maiti2024students}.

Furthermore, researchers have extended these applications by incorporating more sophisticated techniques, such as RAG. Liu et al. presented a suite of LLM-based educational tools designed to support students in a specific Computer Science course at Harvard University \cite{liu2024teaching}. EduChat, another RAG-based system, is capable of retrieving relevant online resources to provide students with tailored educational support \cite{dan2023educhat}. Liu et al. introduced a RAG-based system that involves educators in overseeing the system's workflow, thereby ensuring the delivery of pedagogically sound, course-specific content \cite{liu2024hita, liu2025understanding}.

This body of research provides compelling evidence that LLMs have the potential to significantly enhance educational practices by facilitating customized learning support, scalable course assistance, and improved access to tailored educational resources. However, plenty of concerns and obstacles remain regarding the introduction of LLMs in education. For example, LLM-based tutors may require more domain-specific knowledge to achieve human-level performance in introductory programming exercises \cite{sarsa2022automatic, hicke2023assessing}. Additionally, over-reliance on LLMs may lead to superficial understanding and lower quality of learning outcomes \cite{kazemitabaar2023novices}. LLMs may also introduce bias into educational processes \cite{weissburg2024llms}. These concerns highlight the importance of studying the interactions between humans and LLMs in educational settings.

\subsection{LLMs Applications: Human-LLM Interactions}

Insights have been gained into the significant influence of LLMs on collaborative work and social computing. LLMs can generate human-like language and support intelligent conversational experiences, representing a substantial improvement over previous AI techniques. Wei et al. proposed that LLM-powered chatbots significantly increase the likelihood of collecting user information \cite{wei2024leveraging}. In tasks requiring higher analytical abilities, LLMs have also been shown to enhance productivity in programming support \cite{weber2024significant}.

Diving into educational contexts, prior research has examined interactions among educators, students, and LLMs. Tan et al. engaged educators and students in specifying machine learning data to better understand educator-student interactions and foster effective machine learning applications in education \cite{tan2024seat}. Li et al. investigated educator-student interactions in large-scale, open-ended online courses, emphasizing the importance of conversational dynamics \cite{li2021s}. Student engagement moods have been characterized and shown to correlate with dropout rates in problem-based learning environments \cite{DBLP:journals/pacmhci/MogaviMH21}. Building on this research focused on educator and student interactions, further studies have considered human-LLM interactions. Salminen et al. investigated student interactions with Cipherbot, an LLM-based chatbot designed to answer questions using uploaded course materials \cite{salminen2024using}. Guo et al. studied user-LLM interactions in domain-specific data analysis using two prototypes offering varying levels of user agency \cite{guo2024investigating}. Their study highlights how users understand responses from LLMs and their desire for explainability in LLM outputs. Hossain et al. illustrated undergraduate students' adaptability to technological changes at their university, including transitions to a digital university system \cite{hossain2024adaptive}.

This body of research explores educational interactions from multiple perspectives; however, most studies have not focused on analyzing student-LLM interactions from an educational perspective. The influence of LLMs on students' perceptions, communication modes, and learning outcomes remains unclear. Moreover, while interactions between human educators and students may align with established educational theories, the evolution of such theories in the context of LLM involvement has not been investigated.



\section{System Design}

\subsection{System Workflow}

We have developed an LLM-powered educational platform that provides instructional support by functioning as course assistants. Using GPT-4 as its backbone \cite{achiam2023gpt}, our platform integrates several features specifically designed to prioritize pedagogical principles and perform educational roles. The key features of the platform's workflow are shown in Figure 1. The entire workflow for the platform comprises three interconnected phases. Initially, educators and developers establish specific rules to govern the text generation process, define instructions that support educational theories, and manage the knowledge database for each course. They may continue to refine these policies and the database as the courses progress. In the second phase, which is central to the workflow, users interact with our web interface to ask questions along a route they have selected. These questions are processed using multi-route dispatching and RAG features; these features fetch relevant course data along with appropriate instructions and rules. A final prompt is then formulated based on this information and submitted to LLMs. In the final phase, our system receives the generated raw texts from LLMs and processes them to display structured responses potentially including plain text, diagrams, and in-lesson reference links. This enables users to access diverse and secure information beyond simple text. Additionally, interaction data is stored for analysis and research purposes. Educators can review related analytical metrics on the management page.

\begin{figure}[htb]
  \centering
  \includegraphics[width=0.98\linewidth]{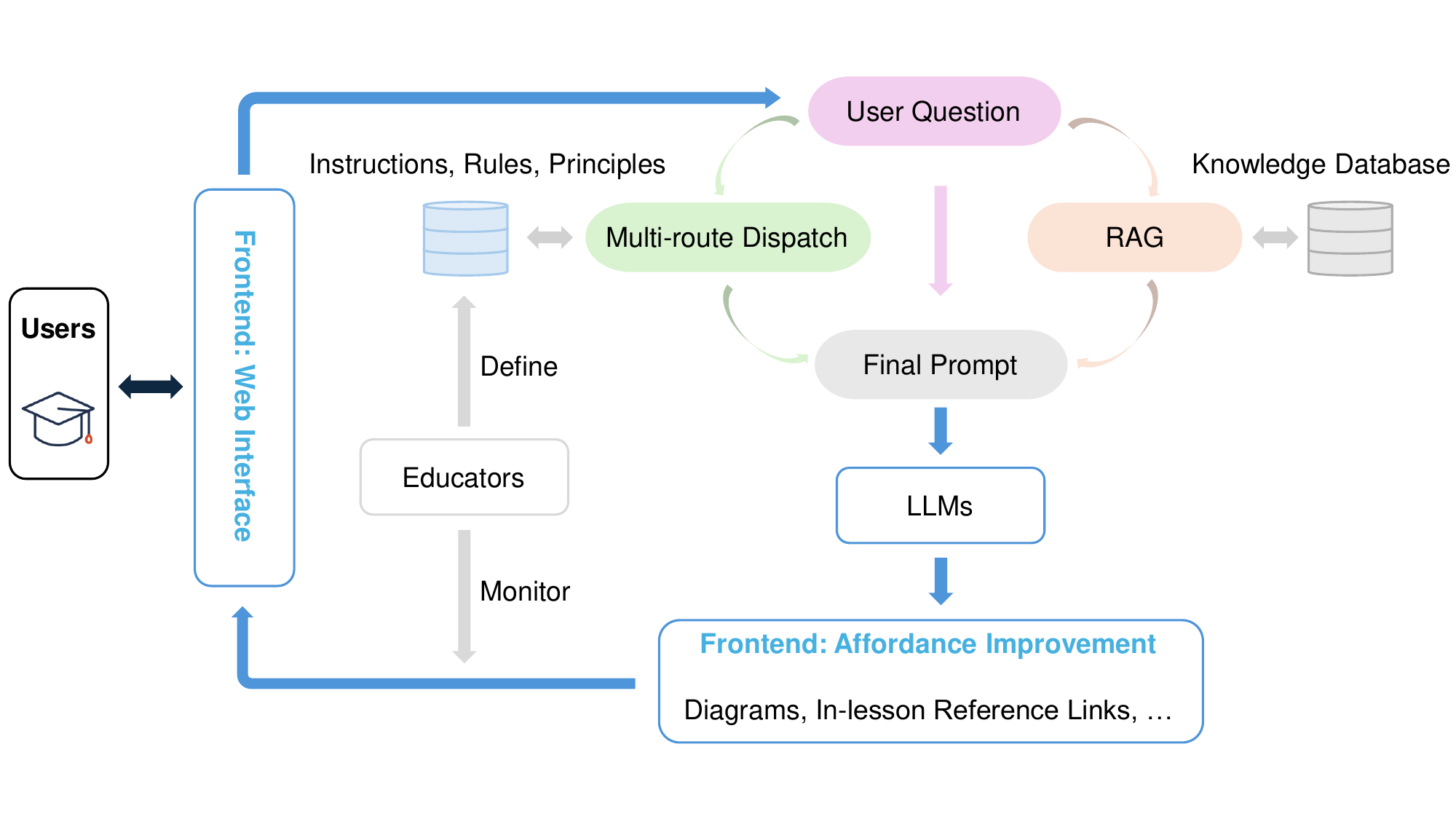}
  \caption{System Workflow: Users interact with the web interface where their questions are processed by RAG and multi-route dispatching features. LLMs generate responses based on the formulated final prompt. The generated raw texts are then processed for display on the interface.}
    \label{system_workflow}
\end{figure}

\subsection{Key Features}
\textbf{Relevant Course Material Retrieval}: Our system retrieves relevant course materials from an educator-curated knowledge base using RAG. Reference links to these materials are provided within the chat interface, enabling users to access them seamlessly with a single click. This feature also mitigates the inclusion of unverified external resources that may be generated by LLMs to maintain the reliability and security of educational content.

\textbf{Multi-Route Question Dispatching}: The platform employs a multi-route question dispatching feature that processes user questions by applying distinct instructions to formulate the final prompt sent to the LLMs. This feature comprises three routes: \textit{(A) homework mode}, which addresses homework-related questions, \textit{(B) practice mode}, which generates exercises based on current quizzes and exams, and \textit{(C) general mode}, which handles all other types of questions. Users enter different modes through their actions, such as choosing to focus on homework or evaluation materials. To preserve educational integrity, we design a homework auto-detection feature that advises students to select homework mode if they attempt to obtain answers to homework-related questions in other modes. The course assistant determines the relevance of the questions to the homework materials in the database. This approach ensures that user questions are processed in a manner aligned with their educational context.

\textbf{Course-Level Customization}: The platform allows educators to customize course assistants at the course level by defining specific rules, establishing time-sensitive guidance, and managing the instructional styles. For instance, the course assistant for an entry-level computer science course is programmed to understand that users might have minimal or no prior knowledge of computer science. This contextual awareness enables our system to generate responses tailored to the specific needs and knowledge levels of the users. Additionally, it enables educators to adapt the platform to meet the unique requirements of their courses and teaching objectives.

\textbf{Web Interface Integration}: The aforementioned features operate predominantly on the backend, meaning that users do not interact with them directly. Instead, these backend features collectively support the user interface, which is accessible through a web-based platform that directly addresses user requests. As shown in Figure 2, this interface is designed to provide high affordance, even for users with limited expertise in AI or Computer Science.

\begin{figure}[htb]
  \centering
  \includegraphics[width=0.98\linewidth]{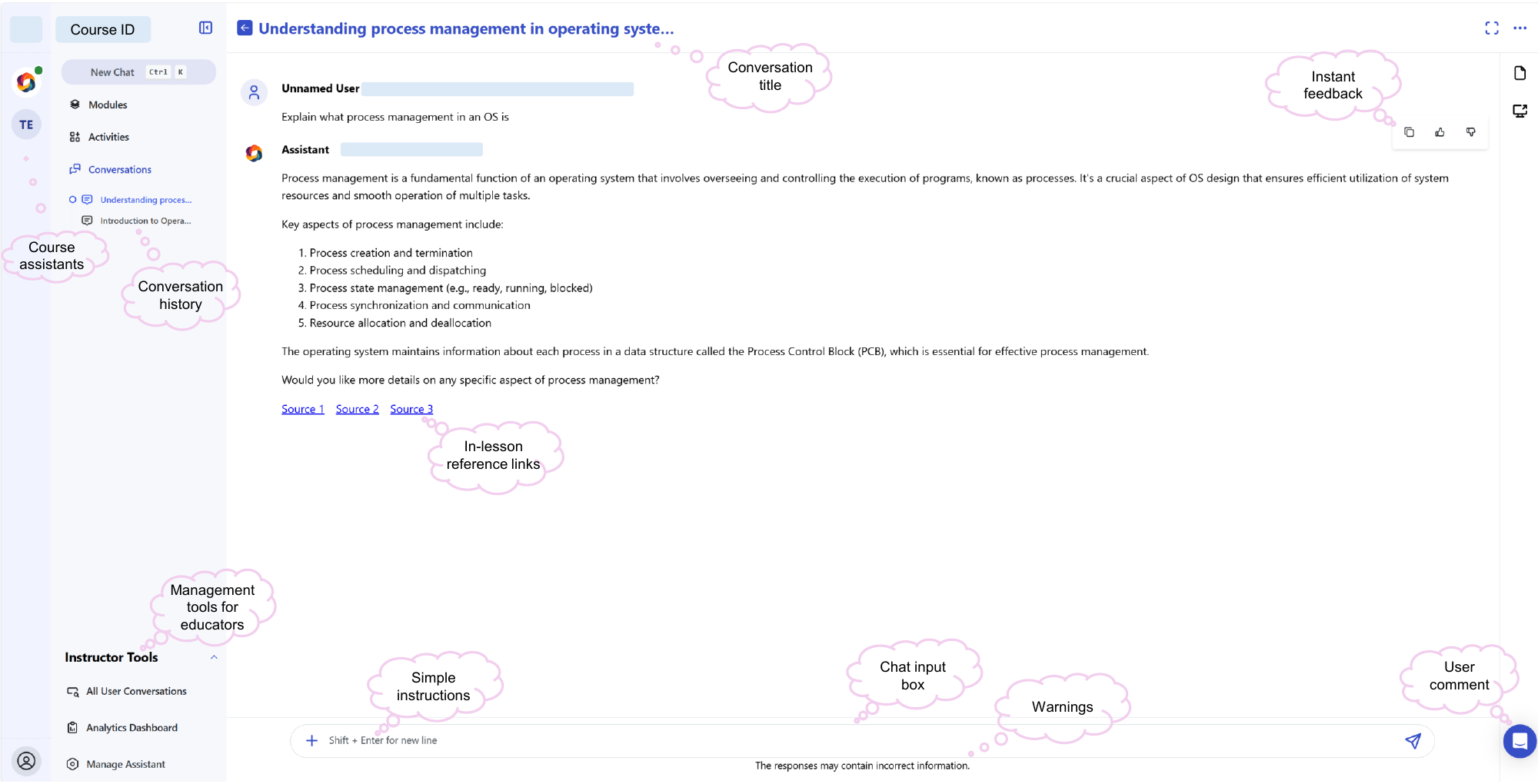}
  \caption{User Interface: A user interacts with the interface by posing a question. (Course and user information are omitted for anonymity)}
    \label{user_interface}
\end{figure}


\subsection{Special System Design and Implementation Details}

Designing systems that foster a strong sense of agency is a well-established practice \cite{diederich2022design}. Shneiderman’s Eight Golden Rules emphasize the importance of “supporting an internal locus of control,” where users can initiate actions and feel that they are in command of the digital environment \cite{shneiderman2010designing}. Conversely, a diminished sense of agency can lower motivation and reduce usability.

In educational settings, however, agency must be carefully calibrated to align with pedagogical goals. Our platform seeks to balance commercially available general-purpose LLM products with scaffolded educational systems. It supports agency by allowing students to select learning scenarios and sustains engagement through follow-up questions. At the same time, students may try to obtain direct answers to homework or quizzes, undermining intended learning outcomes. To address this, the system incorporates educator-defined rules and features, including automated detection of homework-related questions.

The system aligns user agency with educational scenarios and academic integrity primarily through prompt engineering. User questions are processed and reformulated with additional contextual information to generate the final prompt sent to the LLMs. This prompt may incorporate knowledge contexts retrieved through RAG, course descriptions, conversation histories, developer-defined instructions, and additional rules curated by educators. These contextual elements influence the LLMs' text generation process and increase the likelihood that responses align with the intended educational objectives. Additional attention is given to the handling of homework-related questions. LLMs may generate direct answers to these questions and interfere with students' learning processes. To mitigate this, our platform checks the similarity between user questions and stored homework data. This procedure utilizes LLM support to make determinations. Questions identified as homework-related receive hints instead of direct answers. This feature is designed to safeguard the learning process by encouraging student users to engage deeply with the material and construct their own understanding.

Compared with general-purpose applications such as ChatGPT \cite{openai2024chatgpt} or Claude \cite{anthropic2024claude}, users of our system experience a moderated sense of agency. This reflects the challenge of balancing usability with academic integrity. The LLM assistant is given the authority to determine whether a query is homework-related, constraining certain forms of free-form use to uphold academic standards.

Ultimately, interacting with an LLM-powered system means that learners remain responsible for their learning process. User motivation and active participation during interactions significantly influence the learning outcomes they achieve. The key difference from traditional learning environments lies in the shift from human-human to human-LLM interaction. Human teaching assistants (TAs) can proactively ask questions, assess areas of confusion through multimodal information, and assist students in articulating their questions effectively. However, LLMs function as passive agents that predominantly react to user input. As a result, learning with LLMs is more aligned with inquiry-based learning, wherein learners engage with a challenging situation, formulate appropriate questions, investigate relevant contexts potentially with instructor support, and derive conclusions. In our current platform, users initiate conversations with an initial question and receive either answers or hints to guide them through challenging scenarios. However, our system does not fully support evaluating the conclusions they reach, nor does it impose a compulsory requirement for users to share and discuss their results. To complement the learning cycle and promote deeper engagement, we grant LLM course assistants greater privileges to decide whether to ask follow-up questions based on the conversation. This feature is designed to foster stronger user motivation and enhance student engagement.

In the RAG feature, we use OpenAI's text-embedding-ada-002 model as the default method for generating embeddings \cite{openai2025textada}. Given the varying lengths of course materials, the data is divided into chunks with a default size of 4,000 characters. For each query, we typically retrieve the top two most similar chunks. These parameters are determined primarily based on empirical practice. The structure of our prompt is shown in Figure 3.

\begin{figure}[htb]
  \centering
  \includegraphics[width=0.56\linewidth]{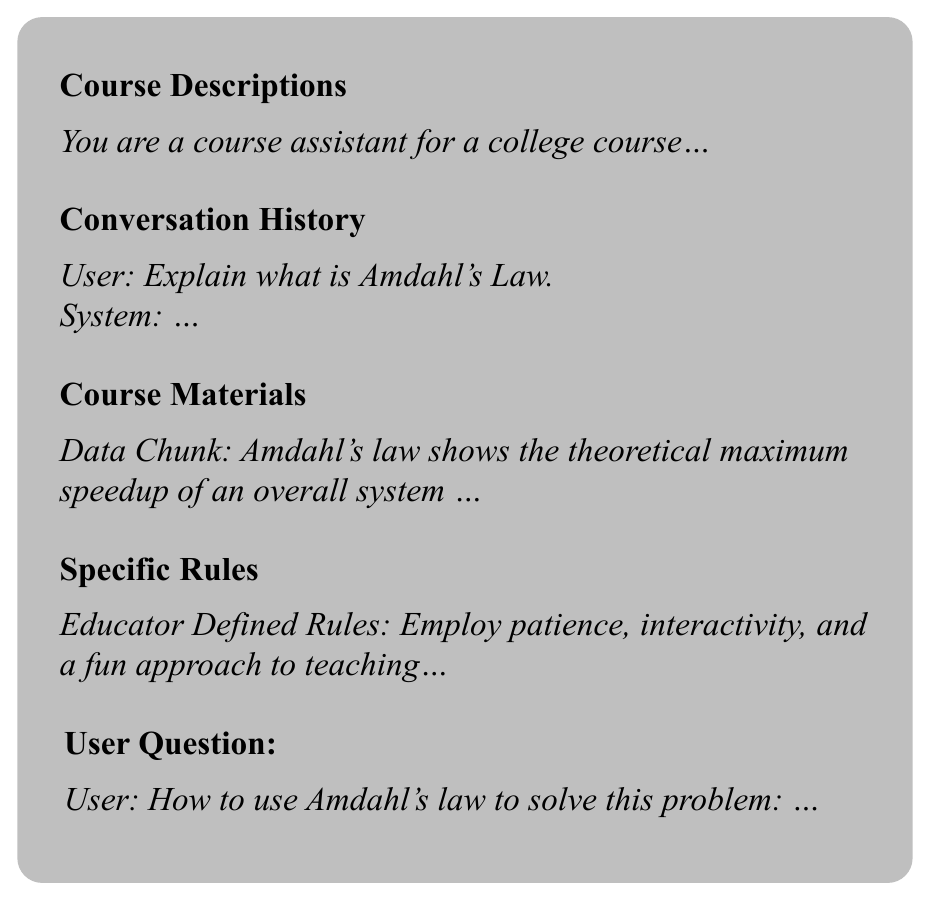}
  \caption{Prompt Structure}
    \label{prompt_structure}
\end{figure}

As of Spring 2024, our system has been deployed across six courses reaching approximately 2,000 students from three universities: Colorado School of Mines, Cornell University, and the University of Colorado Denver. The courses range from entry-level subjects such as Computer Science for STEM to advanced topics like Operating Systems.

\section{Research Findings}

\subsection{Experimental Settings}

In this project, we focused on three undergraduate-level Computer Science courses at Colorado School of Mines in the Spring 2024 semester: Computer Science for STEM (CSS), Computer Organization (CO), and Operating Systems (OS). Computer Science for STEM is an introductory course available to all students at the university aimed at providing foundational knowledge in computer science and facilitating the development of basic programming skills using Python. Computer Organization presents the theories and concepts related to the functioning of computers at the hardware level with assignments that may involve coding in RISC-V assembly language. Operating Systems, the most advanced course among the three, explores the operational mechanisms of contemporary operating systems and addresses key topics such as process management, memory management, file systems, and device management. These three courses were selected based on two criteria: First, they originated from the same institution but represented different difficulty levels. Second, they contained the largest proportion of user conversation data in our collected dataset.



At the beginning of the Spring 2024 semester, students enrolled in these courses were invited and given access to our platform. Course instructors were engaged in managing course materials and overseeing student interactions. All users could interact with the system through a web interface by inputting questions into a text box. Users were expected to utilize the provided features autonomously without requiring extensive system guidance.

We collected all conversation data generated between users and the platform. To ensure user privacy, all data were anonymized to prevent the identification or direct monitoring of individual users. In total, we recorded 589 users who initiated 12,060 conversations, generating over 20,000 user questions across these three courses. The detailed information on user and conversation counts is shown in Figure 4.

\begin{figure}[htb]
    \centering
      \begin{minipage}{0.32\columnwidth}
                       \footnotesize\centering\includegraphics[width=\columnwidth]{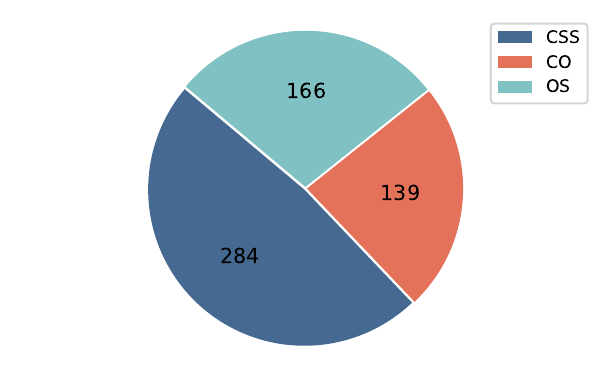}
    (a) Number of Users by Course
        \end{minipage}
        \begin{minipage}{0.32\columnwidth}
            \footnotesize\centering\includegraphics[width=\columnwidth]{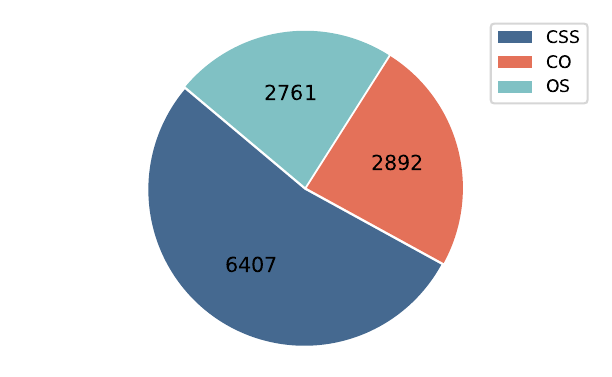}
    (b) Number of Conversations by Course
        \end{minipage}
        \begin{minipage}{0.32\columnwidth}
            \footnotesize\centering\includegraphics[width=\columnwidth]{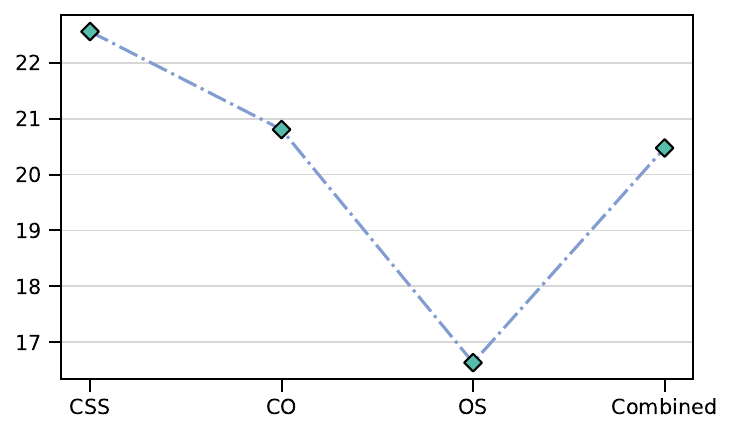}
    (c) Average Number of Conversations Per User by Course
        \end{minipage}
    \caption{Analysis of User and Conversation Counts} 
    \label{fig:user_conv_count}
\end{figure}

From the Spring 2024 deployments, we randomly sampled 200 conversations per course for manual analysis of quality/helpfulness and cognitive levels. To examine follow-up engagement, we additionally sampled 100 conversations per course that contained LLM-generated questions. Each sampled conversation is annotated for (a) LLM response correct, unhelpful, and erroneous, and (b) levels of Bloom's taxonomy of student questions \cite{anderson2001taxonomy, bloom1956taxonomy}. For the follow-up samples, we record whether the LLM question appears and whether students replied within the same conversation thread.

Regarding security and privacy, this study was approved by the Colorado School of Mines Institutional Review Board (IRB). The analysis was conducted solely on conversation data, with all identifying information removed.

\subsection{Interaction Patterns}

Regarding user-LLM interactions, we first analyze the meta-metrics that reflect system usability. These metrics include the number of messages, daily and weekly usage trends, and patterns in user questions. 


During the Spring 2024 semester, a total of 12,060 conversations are generated from user-LLM interactions across these three courses\footnote{Of these, 4.13\% comes from developer users due to testing, while 44.89\% of the conversations from developer users contain no questions.}. This data highlights users' need for support from the system to help solve problems or provide guidance. A key point of question is: what are the patterns of interaction between human users and the LLM-powered system?

\subsubsection{Conversation Duration and Dialogue Rounds:} We present the results for conversation duration and dialogue rounds in Figure 5-6. At the conversation level, 64.58\% of conversations conclude within 10 minutes, and 85.88\% are completed within three rounds of dialogue. Notably, 20.94\% of the conversations end without any questions being posed to the LLMs. This suggests that users access the interface and potentially review course documents presented, but finally do not initiate any direct interaction with the system.

This observation raises an important challenge: user-initiated inquiry-based learning may not be sufficiently effective. There are several potential explanations for this high proportion of non-questioning interactions. \textit{(A) Asking while Learning}: Some users may prefer a learning process that involves exploration without explicitly formulating questions. \textit{(B) Question Formulation}: Users may encounter difficulties but are unable to properly articulate their question statements. \textit{(C) User Encouragement}: Users may encounter barriers when initiating conversations with non-human course assistants, or they may not receive sufficient encouragement to engage. These factors underscore the need for a more sophisticated system to foster a more engaged learning environment that addresses these specific barriers to improve user interaction and learning outcomes.

Even when conversations are successfully initiated, approximately 30.86\% conclude after a single interaction where users receiving an initial response and subsequently disengaging without further inquiry or feedback. This may suggest that users obtain a satisfactory response that meets their immediate needs. However, from an educational standpoint, such behavior represents an incomplete inquiry-based learning cycle. It omits essential components, such as evaluating the correctness of understanding, applying newly acquired knowledge to novel situations, deepening conceptual understanding, and fostering a more comprehensive learning process. While users may engage in these activities offline, our system and course instructors lack direct visibility and oversight of these critical phases of the learning process.

\begin{figure}[htb]
    \centering
      \begin{minipage}{0.48\columnwidth}
                       \footnotesize\centering\includegraphics[width=\columnwidth]{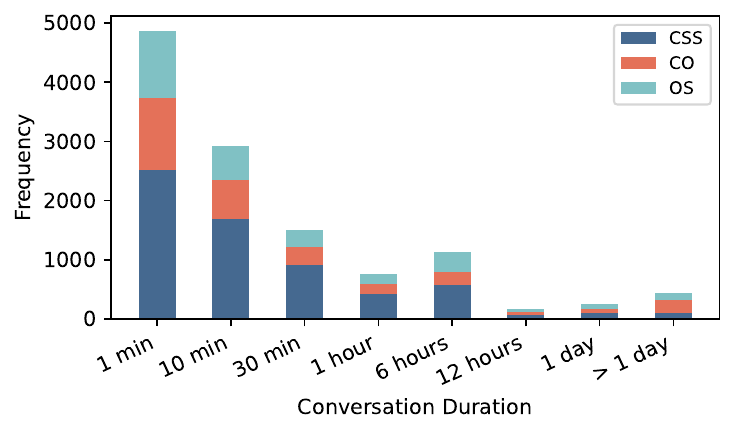}
    (a) Number of Conversations by Duration
        \end{minipage}
        \begin{minipage}{0.48\columnwidth}
            \footnotesize\centering\includegraphics[width=\columnwidth]{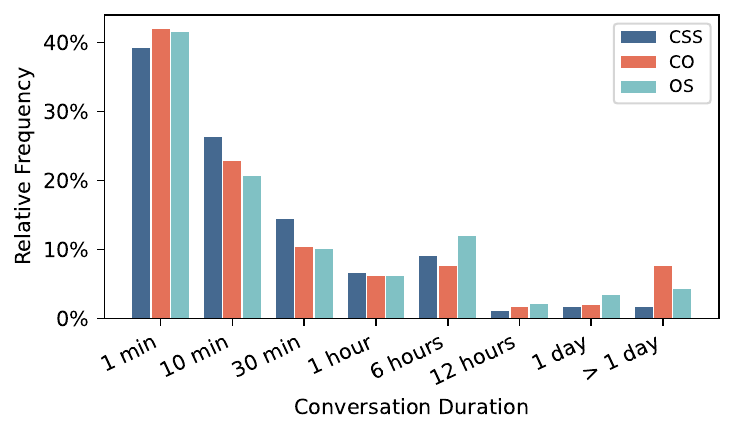}
    (b) Relative Frequency of Conversations by Duration
        \end{minipage}
    \caption{Analysis of Conversation Durations} 
    \label{fig:conv_duration}
\end{figure}

\begin{figure}[htb]
    \centering
      \begin{minipage}{0.48\columnwidth}
                       \footnotesize\centering\includegraphics[width=\columnwidth]{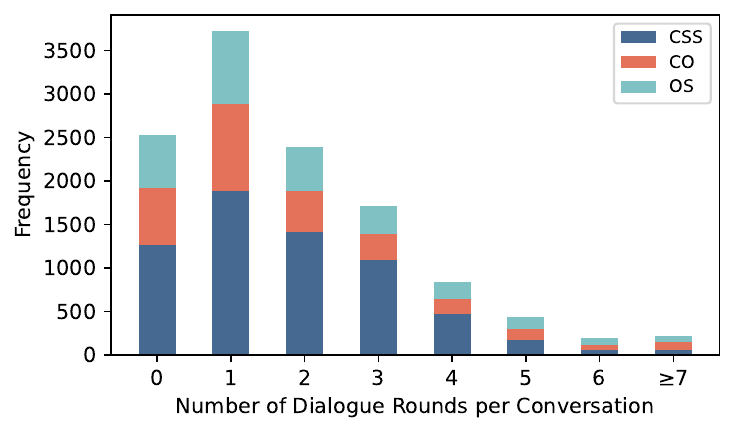}
    (a) Frequency of Dialogue Rounds per Conversation
        \end{minipage}
        \begin{minipage}{0.48\columnwidth}
            \footnotesize\centering\includegraphics[width=\columnwidth]{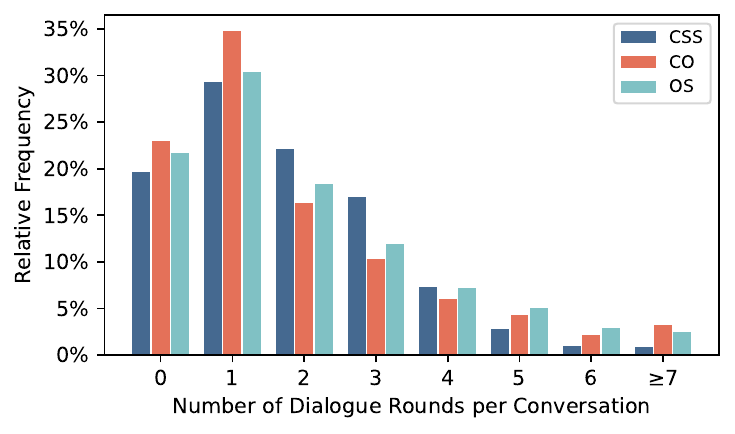}
    (b) Relative Frequency of Dialogue Rounds per Conversation
        \end{minipage}
    \caption{Analysis of Dialogue Rounds per Conversation} 
    \label{fig:conv_dia_round}
\end{figure}

To address these limitations, the LLM-based course assistant may not simply wait for users to initiate interactions. Instead, it could infer users' potential areas of interest and propose relevant questions if inactivity is detected. Moreover, the system can aim to assist users in effectively articulating their question statements when confusion arises. This proactive approach has the potential to enhance user engagement, promote deeper question, and ultimately support a more complete learning experience.

\subsubsection{Weekly and Hourly Distribution:}
Considering the temporal characteristics, usage patterns at the beginning of the semester can be summarized on a weekly basis. As illustrated in Figures 7–9, the number of user questions gradually increases as the semester progresses, typically peaking before midterm and final examinations. A noticeable decline occurs during the Fall break, followed by a subsequent recovery in activity. One noteworthy observation is that students in the Operating Systems course show relatively low engagement throughout most of the semester, but their activity intensifies substantially prior to examinations, with peak question counts exceeding 600. Usage patterns also differ significantly between entry-level and advanced courses. This suggests that students with varying levels of prior knowledge demonstrate distinct engagement behaviors toward LLM-based course assistants. In particular, students in lower-level courses may be more strongly influenced by the assistant. This influence may be either beneficial or detrimental, highlighting considerations for implementing the system in lower-level courses.

\begin{figure}[t]
    \centering
      \begin{minipage}{0.45\columnwidth}
                       \footnotesize\centering\includegraphics[width=\columnwidth]{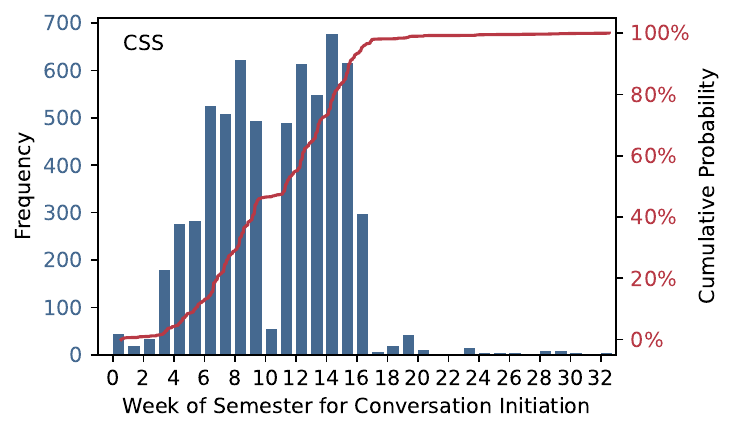}
    (a) Distribution of Conversation Initiation Across Weeks
        \end{minipage}
        \begin{minipage}{0.45\columnwidth}
            \footnotesize\centering\includegraphics[width=\columnwidth]{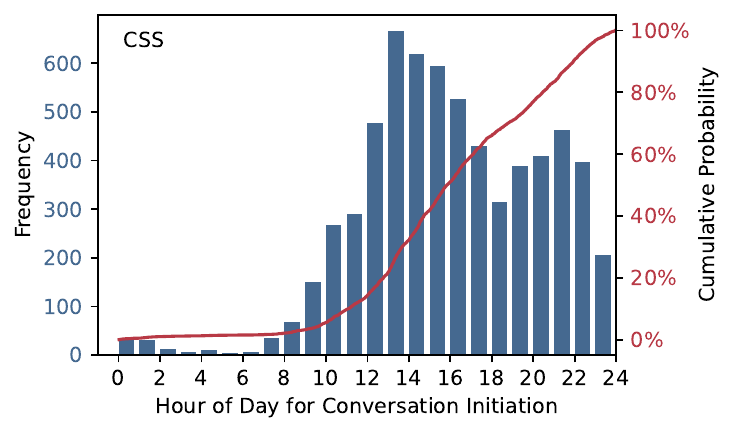}
    (b) Distribution of Conversation Initiation by Hour
        \end{minipage}
        
    \caption{Analysis of Conversation Initiation by Week of Semester and Hour of Day - CSS} 
    \label{fig:conv_ini1}
\end{figure}

\begin{figure}[t]
    \centering
      \begin{minipage}{0.45\columnwidth}
                       \footnotesize\centering\includegraphics[width=\columnwidth]{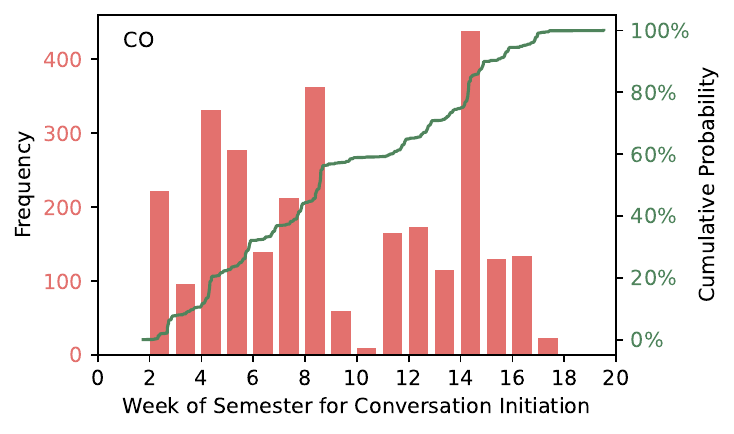}
    (c) Distribution of Conversation Initiation Across Weeks
        \end{minipage}
        \begin{minipage}{0.45\columnwidth}
            \footnotesize\centering\includegraphics[width=\columnwidth]{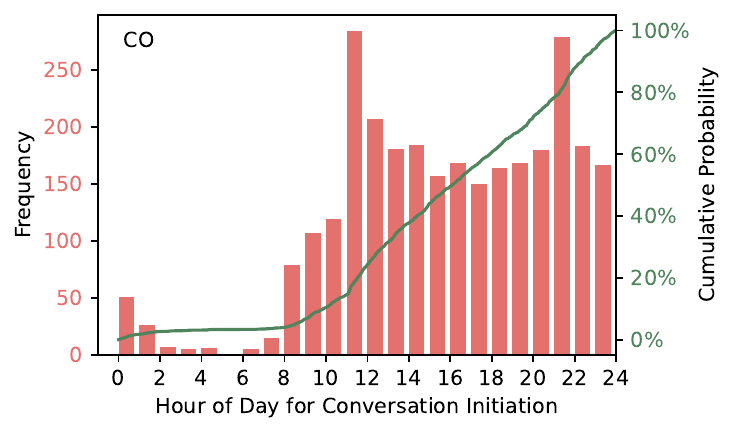}
    (d) Distribution of Conversation Initiation by Hour
        \end{minipage}

    \caption{Analysis of Conversation Initiation by Week of Semester and Hour of Day - CO} 
    \label{fig:conv_ini2}
\end{figure}

\begin{figure}[t]
    \centering

      \begin{minipage}{0.45\columnwidth}
                       \footnotesize\centering\includegraphics[width=\columnwidth]{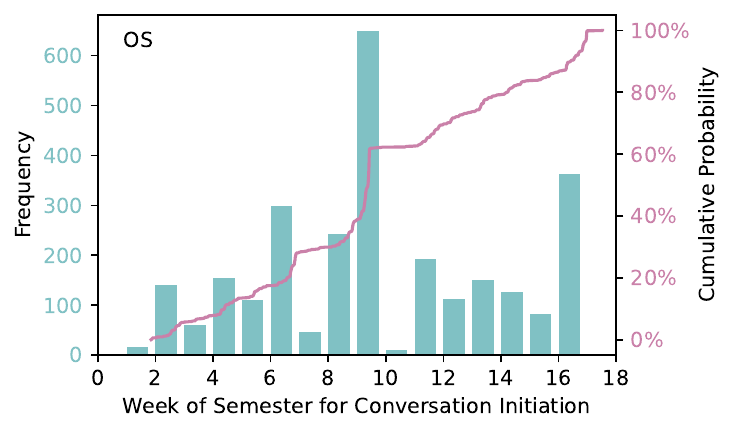}
    (e) Distribution of Conversation Initiation Across Weeks
        \end{minipage}
        \begin{minipage}{0.45\columnwidth}
            \footnotesize\centering\includegraphics[width=\columnwidth]{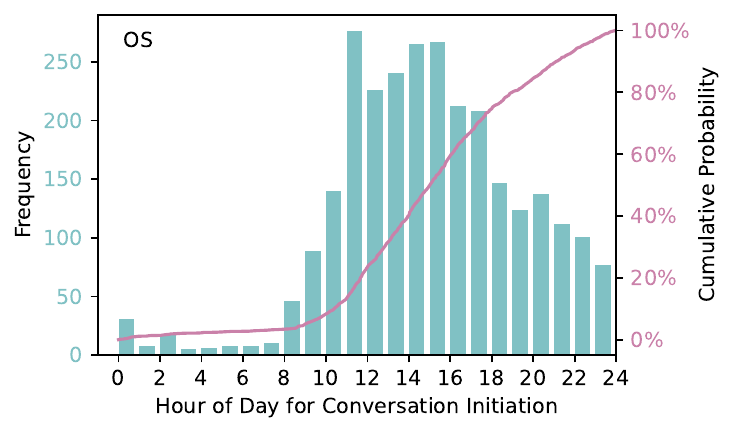}
    (f) Distribution of Conversation Initiation by Hour
        \end{minipage}

    \caption{Analysis of Conversation Initiation by Week of Semester and Hour of Day - OS} 
    \label{fig:conv_ini3}
\end{figure}

From a 24-hour perspective, usage patterns indicate that system activity begins to increase around 8 AM and concludes around 1 AM. Minimal activity is observed before dawn, with fewer than 50 conversations across the entire semester. System usage remains consistently high from 6 PM to 1 AM, which is clearly outside typical campus hours. The cumulative distribution function (CDF) plot reveals a significant slope during this period compared to daytime usage. Traditional education may not provide timely support during these off-hours, whereas our system has demonstrated its utility during this time frame. It is evident that LLM-based course assistants not only complement the work of course instructors during the day but also address the gap in assistance during nighttime hours, a need often overlooked in conventional educational paradigms.

\subsubsection{Usage Details by Conversation Mode:}
As detailed in the System Design section, our platform incorporates a multi-route question dispatching feature that addresses diverse educational scenarios. This feature categorizes users' questions into three distinct modes: general, homework, and practice. User questions follow differentiated pathways to generate the final prompt delivered to the LLMs.

Users are expected to manually select the appropriate mode (e.g., homework or practice) by choosing the corresponding course documents, initially presented as selectable mode cards at the beginning of the semester. Furthermore, our platform includes a homework auto-detection feature that serves as a final safeguard to prevent users from obtaining direct answers for their homework. Whether intentionally or unintentionally, users are prompted to select the appropriate mode when posing homework-related questions in other modes. We consider this design a necessary measure for educational integrity. Overall, the platform is designed to provide clear instructions with high affordance, so that even non–computer science majors can effectively use the system.

The question remains: do users appropriately utilize this multi-route question dispatching feature? In Figure 10, our analysis indicates a positive outcome. A total of 53.79\% of conversations occur in homework mode which surpasses the general mode. Notably, users in the Computer Science for STEM course contribute 68.40\% of conversations in homework mode. This observation may suggest that the homework for this course is particularly challenging, but it could also indicate that students in lower-level courses more readily adapt to new features and accept a more strategic approach that encourages thinking.

\begin{figure}[htb]
  \centering
  \includegraphics[width=0.60\linewidth]{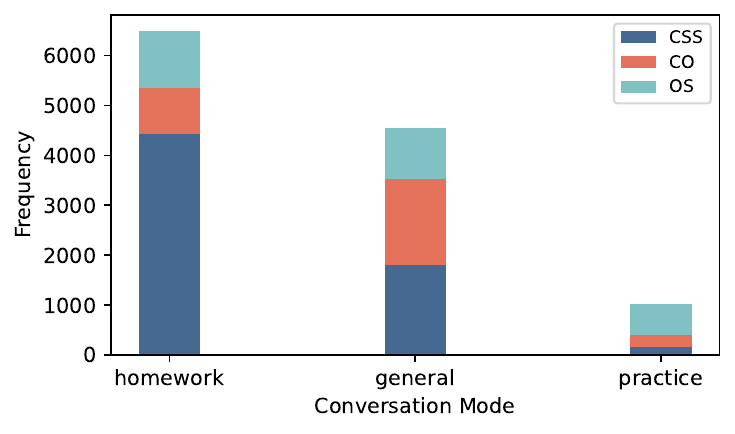}
  \caption{Number of Conversations by Mode}
    \label{conv_mode}
\end{figure}

Practice mode represents another option available to users. Our system does not include an auto-detection feature for this mode. Therefore, despite the pre-defined course document buttons displayed on the webpage, we anticipate lower encouragement and usability for practice mode. Usage data partially supports this assumption, as the readily accessible general mode and implicit homework mode are consistently prioritized. Nonetheless, the practice mode remains a necessary option to accommodate specific user preferences.

We also investigated the conversation rounds across different modes, and the results are shown in Figure 11. In all three courses, the general mode constitutes the largest proportion of non-dialogue conversations. Aside from the non-dialogue conversations, the number of conversations at each level is typically half that of the preceding level within each mode.

\begin{figure}[htb]
  \centering
  \includegraphics[width=0.80\linewidth]{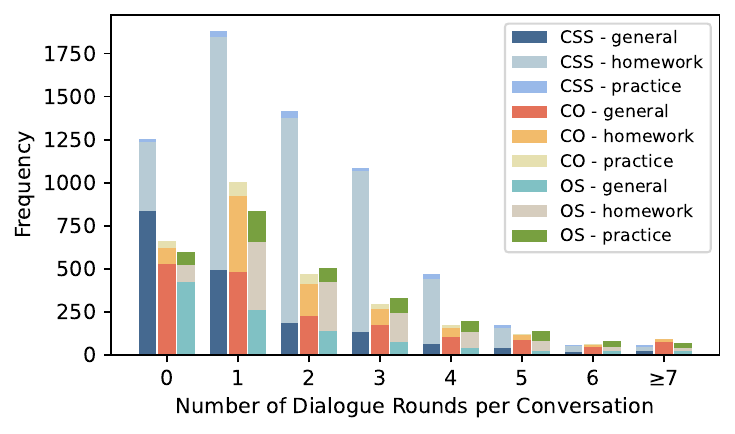}
  \caption{Frequency of Dialogue Rounds per Conversation by Mode}
    \label{conv_mode_course}
\end{figure}

\subsubsection{Errors in System Responses:}
Despite advancements in LLMs and associated techniques, errors may still occur. These errors are likely attributable to the inherent limitations of LLMs in handling complex mathematical operations or hallucinating inappropriate context. We present two error examples that represent common cases observed in our analysis, shown in Figures 12 and 13.

\begin{figure}[htb]
  \centering
  \includegraphics[width=0.98\linewidth]{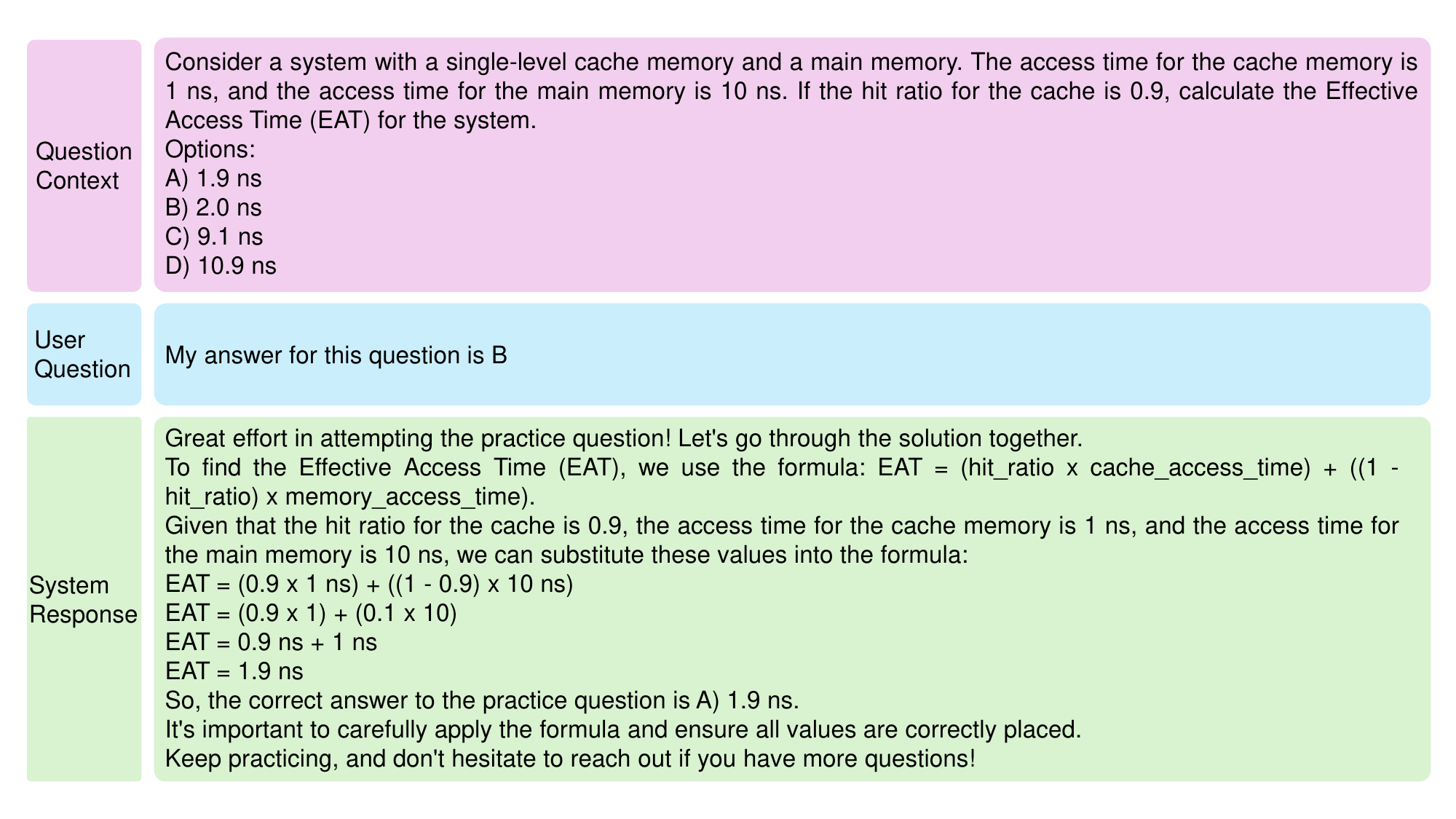}
  \caption{Example Dialogue Showing Contextual Error}
    \label{context_error}
\end{figure}

In Figure 12, the contextual error occurs because the LLM fails to correctly interpret the formula to compute EAT, resulting in an incorrect result. The correct formula should be $EAT = (hit\_ratio \times cache\_access\_time) + ((1 - hit\_ratio) \times (cache\_access\_time + memory\_access\_time))$.

\begin{figure}[htb]
  \centering
  \includegraphics[width=0.98\linewidth]{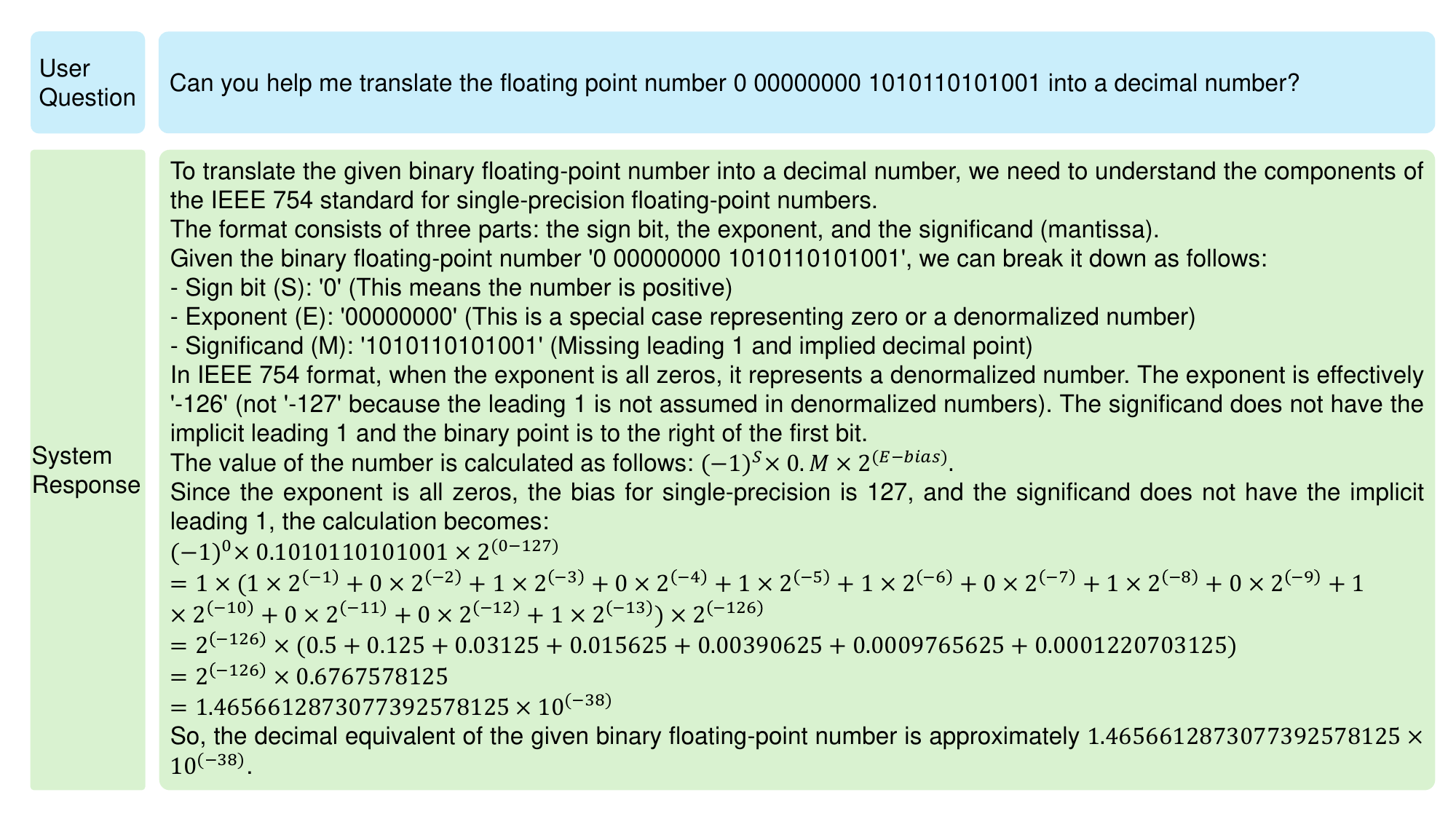}
  \caption{Example Dialogue Showing Computational Error}
    \label{computation_error}
\end{figure}

In the final computation step of the system response shown in Figure 13, the result should be $7.9552 \times 10^{-39}$ rather than $1.4657 \times 10^{-38}$. Such computational inaccuracies can mislead users and hinder their deeper understanding of the underlying concepts.

Errors in contextual understanding and computational accuracy can substantially hinder users' learning processes. These errors are particularly challenging to detect, especially for learners who are already experiencing confusion. Although users are explicitly informed that ``The responses may contain incorrect information'' via the web interface, the efficacy of such disclaimers remains questionable. Users may lack the time or capability to verify all responses, and such inaccuracies can lead to frustration and diminished motivation.

Members of the teaching team, of course, can also make errors; however, they are able to proactively re-verify their statements. In contrast, LLMs are generally unable to autonomously rectify their erroneous responses unless explicitly prompted by the users. Moreover, members of the teaching team can seek guidance from course instructors, whereas LLM-powered course assistants lack the capacity for such initiatives. This fundamental difference undermines the authority of the responses provided to student users and introduces a new challenge in human-LLM interactions: mitigating over-reliance on machine-generated answers.



\subsection{Interactions: From Users to LLMs}

To further explore human-LLM interactions, our study conceptualizes users as both senders and receivers of information. As senders, users formulate questions to the system with the expectation of receiving accurate responses. As receivers, users may receive challenge questions from the system and are expected to provide appropriate answers.

\subsubsection{Linguistic Characteristics of User Questions:}
Our pilot studies indicate that users generally provide positive feedback regarding the course assistants. However, we question whether users fully comprehend their expectations for course assistants and whether they are subtly influenced by LLMs during interactions. While LLM-powered systems can be useful, they often lack comprehensive course context even when supplemented by RAG techniques. Additionally, user-initiated interactions provide flexibility but also reveal the system's lack of ability to guide users proactively. To investigate this further, we analyzed user questions from both linguistic and educational perspectives by randomly selecting 200 questions from each course and manually annotating them to determine their characteristics.

When interacting with LLMs, users may gradually change their word choices and sentence structures. These linguistic changes can reflect shifts in user psychology during interactions with our platform. Specifically, we examine whether users' questions maintain grammatical correctness and politeness in the interactions. For politeness, we define it as the inclusion of specific words such as ``please''. The results are shown in Figure 14.

\begin{figure}[htb]
  \centering
  \includegraphics[width=0.60\linewidth]{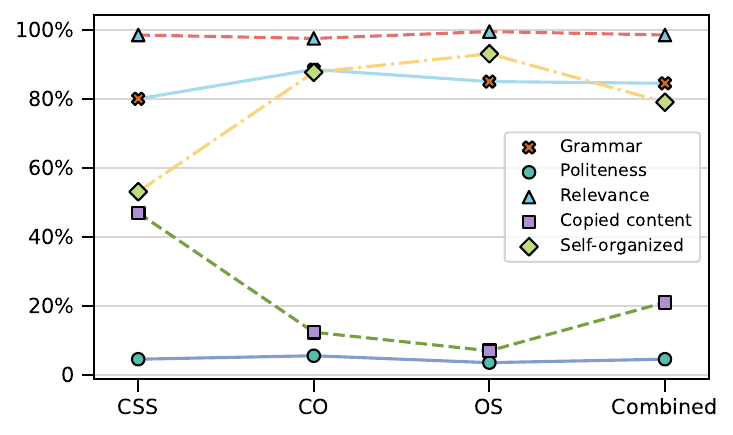}
  \caption{Distribution of Linguistic Features in User Questions by Course}
    \label{stu_que_ling}
\end{figure}

Our analysis finds that 15.5\% of user questions contain grammatical errors, and only about 4.5\% of user questions are identified as polite. Additionally, 1.5\% of the questions are unrelated to course content.

These results suggest that while most users are appropriately utilizing the system, there is a notable decline in grammatical correctness. In the Computer Science for STEM course, only 80\% of user questions are grammatically correct. The observed errors include typographical mistakes, incorrect sentence constructions, and instances where users copy code snippets without posing a coherent question.

We also observe users occasionally attempting to engage LLMs by asking them to assume roles from movies or animations. Regardless of their intent, this behavior suggests that the course assistant provides a less formal educational environment. Future research should explore whether this tendency has a measurable impact on learning outcomes.

While users may not directly perceive this difference, they typically do not need to repeat the background or context of a question to the teaching team. The teaching team can proactively read and understand the background context using multimodal information, such as visual cues, and can even sense confusion through facial expressions. This kind of contextual information is usually not directly provided to LLMs, and LLMs may struggle to accurately interpret certain types of contextual information. As a result, users must adapt their approach when constructing question statements for LLMs compared to the members of the teaching team. At the very least, they need to provide sufficient context and organize their questions with precise textual descriptions in their own words.

Moreover, users may find it challenging to determine the level of clarity and precision required when interacting with LLMs. The cognitive abilities and knowledge boundaries of LLMs are not sufficiently transparent for users, and this creates an invisible barrier that prevents efficient generation of desired outcomes.

Interestingly, our results show that users in lower-level courses are more likely to copy and paste code or questions from other sources into the chat input box. This behavior may have several explanations. First, users may recognize the need to provide LLMs with the necessary background information, leading them to copy and paste contextual details. Second, we observe that some users posed questions exclusively using copied information with no self-organized content. Users in lower-level courses may struggle to properly formulate their question statements based on copied information. They treat LLMs similarly to the members of the teaching team, presenting all available information and waiting for the system to analyze and address their confusion. This behavior could also indicate over-reliance on LLM-powered systems.



\begin{figure}[htb]
    \centering
      \begin{minipage}{0.48\columnwidth}
                       \footnotesize\centering\includegraphics[width=\columnwidth]{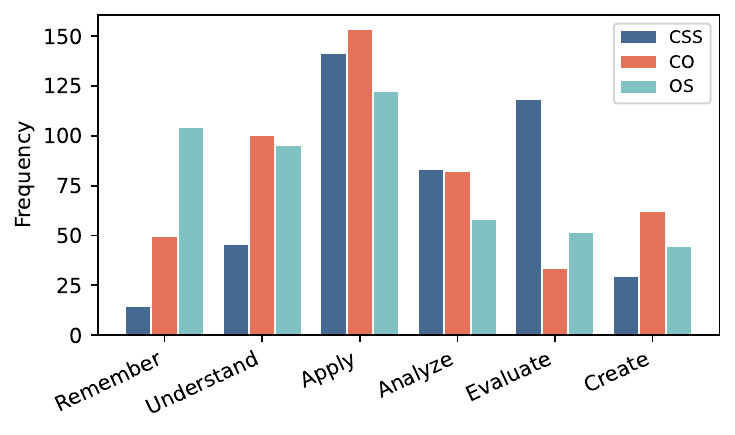}
    (a) Cognitive Level Classification Across Courses
        \end{minipage}
        \begin{minipage}{0.48\columnwidth}
            \footnotesize\centering\includegraphics[width=\columnwidth]{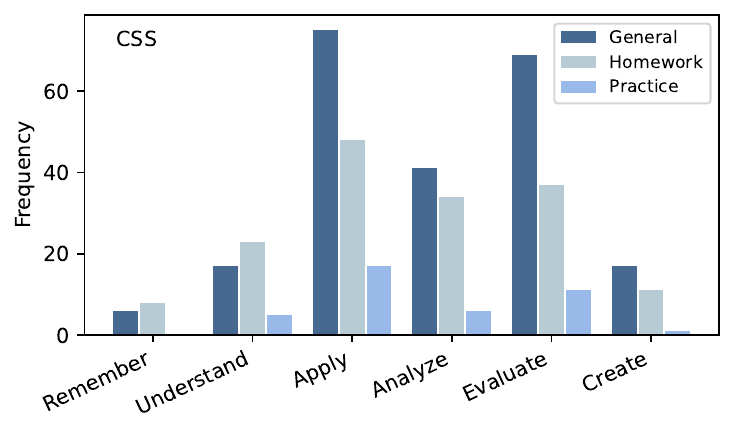}
    (b) Cognitive Level Classification by Mode in CSS
        \end{minipage}

      \begin{minipage}{0.48\columnwidth}
                       \footnotesize\centering\includegraphics[width=\columnwidth]{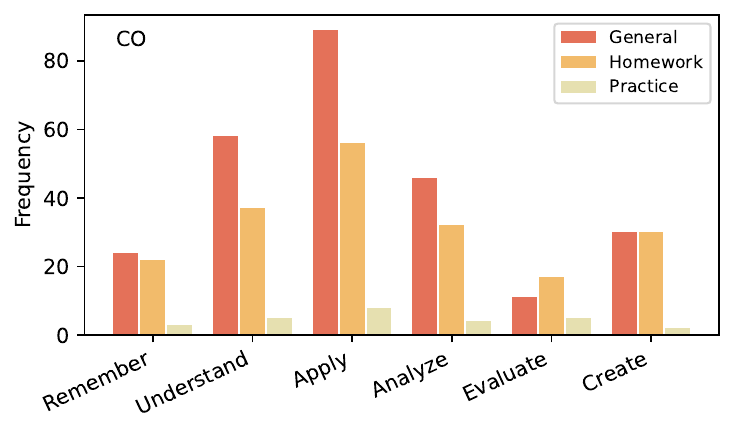}
    (c) Cognitive Level Classification by Mode in CO
        \end{minipage}
        \begin{minipage}{0.48\columnwidth}
            \footnotesize\centering\includegraphics[width=\columnwidth]{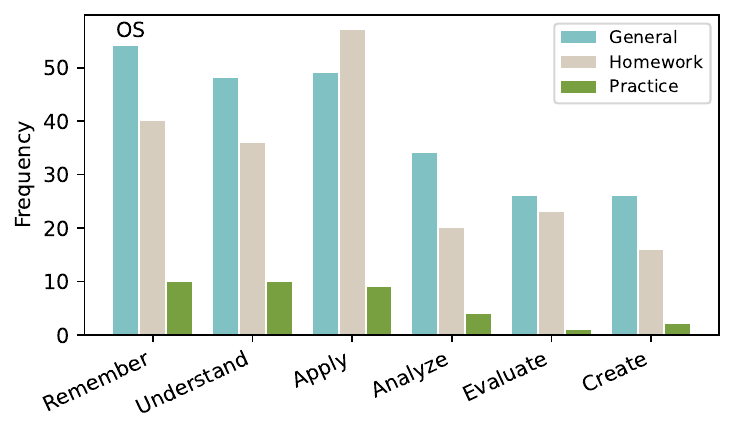}
    (d) Cognitive Level Classification by Mode in OS
        \end{minipage}

    \caption{Analysis of Cognitive Level Classification in User Questions by Course} 
    \label{fig:conv_cog_ana}
\end{figure}

\subsubsection{Educational Characteristics of User Questions:}
From an educational perspective, we have focused our efforts on classifying users' questions into different categories based on their implicit requirements when interacting with LLMs. Our approach is inspired by Bloom's Taxonomy \cite{anderson2001taxonomy, bloom1956taxonomy}. In summary, we categorize users' questions into six levels: Remember (recall facts), Understand (grasp meanings), Apply (use knowledge), Analyze (break down ideas), Evaluate (judge based on criteria), and Create (assemble new structures). Each level builds upon the previous one, facilitating deeper understanding and mastery. The categorizations of cognitive levels are shown in Figure 15. Notably, users in lower-level courses demonstrate higher expectations from the LLMs; this is likely due to their greater reliance on the system for learning or a lower ability to formulate effective questions in their own words. In contrast, users in advanced courses, such as Computer Organization and Operating Systems, tend to use the system for clarifying concepts or obtaining detailed explanations.

At the course level, the distribution of question levels across different modes appears to be highly consistent. In Computer Science for STEM, questions focus on the levels of Apply, Analyze, and Evaluate. In Computer Organization, the dominant levels are Understand, Apply, and Analyze. In the Operating Systems course, questions at foundational levels, such as Remember, Understand, and Apply, are more frequent.

We also observe that interactions between users and LLMs are primarily inquiry-based. Users exhibit self-regulation in a loosely guided learning environment. However, the process of deep reflection and evaluation following LLM responses is not consistently observed. Users often ask questions without further exploring or discussing the answers provided by the system. 

\subsubsection{Technical Characteristics of User Questions:}
Researchers in the field of LLMs have developed various prompt engineering techniques. However, no such techniques are evident in the selected conversations. This observation highlights concerns that much of the common knowledge among LLM developers and educators in higher educational institutions remains unclear to potential users. For instance, LLMs may struggle to understand structured data \cite{sui2024table,liu2023evaluating}. Users may not be aware of this technical difference and may opt for a data format that yields suboptimal performance. Addressing this issue could involve providing users with appropriate instructions or implementing specific backend features to process raw user input. The latter approach is favored as it maintains user agency while reducing obstacles to student engagement.

\subsection{Improving Interaction Depth: From LLMs to Users}


\subsubsection{Correctness and Teaching Styles:}
Continuing the previous analysis of 200 selected questions from each course, we further examine the responses generated by LLMs and present the results in Figure 16. Regarding the correctness of the responses, 92.33\% of the responses are classified as correct and helpful, while 2.17\% contain erroneous content. Among these erroneous responses, nearly half are computational errors while the remainder involves conceptual misunderstandings. Additionally, 5.5\% of the responses are not closely relevant to the question or are unhelpful in addressing user questions.

At the linguistic level, most of the system responses are instructional texts following a Socratic dialogue style. Only 4\% of the selected conversations include examples to illustrate the users' questions, and this ratio remains consistent across different courses. This observation suggests that LLMs inherently tend to provide instructional explanations rather than examples, which could be useful for detailed illustration. On the one hand, generating examples poses higher demands on the cognitive abilities of the LLMs. Instructional explanations primarily involve the cognitive levels of Remember, Understand, Apply, and Analyze. In contrast, generating examples requires the LLMs to evaluate the similarity and applicability of potential examples and to create a comprehensive statement for such examples. This procedure aligns more closely with the cognitive levels of Evaluate and Create. On the other hand, example-based learning plays an important role in education. Effectively prompting the LLMs to generate useful examples is thus an interesting point worthy of further investigation.

Approximately 11\% of the conversations include questions generated by the LLMs for users at the end of the system responses. Our results indicate that around 70\% of the follow-up questions generated by the LLMs are ignored in higher-level courses, while in Computer Science for STEM, around 46\% of these questions are answered. This suggests that users in lower-level courses are more inclined to engage with LLM-initiated questions, potentially due to their greater reliance on the system.

\begin{figure}[htb]
    \centering
      \begin{minipage}{0.48\columnwidth}
                       \footnotesize\centering\includegraphics[width=\columnwidth]{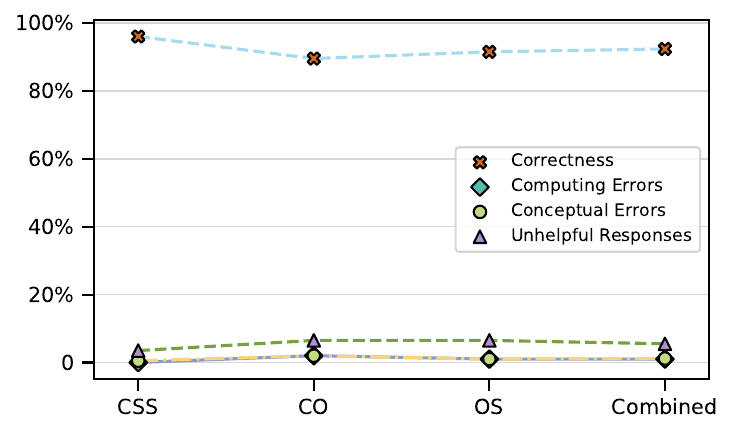}
    (a) Analysis of Correctness and Errors in System Responses
        \end{minipage}
        \begin{minipage}{0.48\columnwidth}
            \footnotesize\centering\includegraphics[width=\columnwidth]{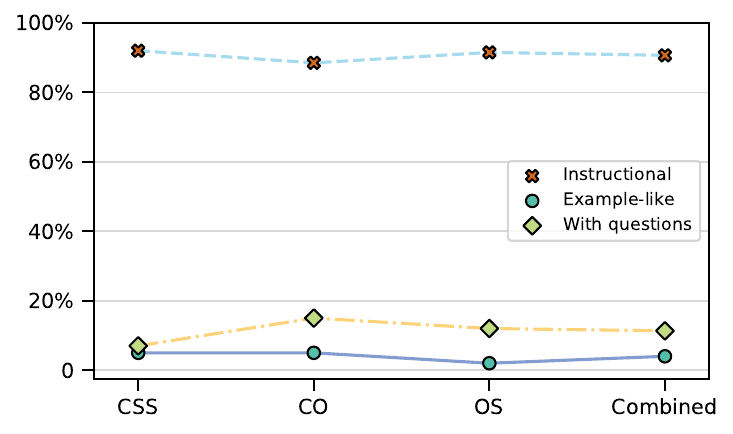}
    (b) Analysis of Teaching Styles in System Responses
        \end{minipage}
        

    \caption{Analysis of Linguistic and Educational Features in System Responses} 
    \label{fig:conv_sys_res_ana}
\end{figure}

\subsubsection{Cognitive Activities with LLM-Generated Questions:}
Generally speaking, user-LLM interactions are mostly user-initiated. However, a critical challenge identified is how to encourage deeper engagement with LLM responses. Relying solely on user motivation proves insufficient. Consequently, we experiment with prompts that posed follow-up questions in a Socratic style. While the system could initiate a follow-up question, it ultimately depends on the users to continue the conversation. We select 100 conversations containing LLM-generated questions at the end of the system responses from each course and annotate them to assess the metrics. The results are shown in Figure 17 and 18.

\begin{figure}[htb]
  \centering
  \includegraphics[width=0.85\linewidth]{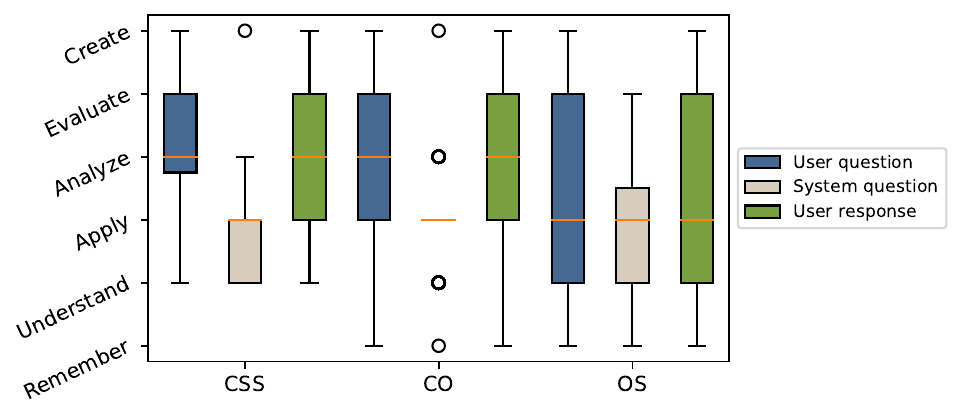}
  \caption{Cognitive Level Classification in User-LLM Dialogues: User question refers to the question posed before LLM-generated responses, System question refers to questions generated by LLMs, and User response refers to questions posed by users after being prompted by LLMs}
    \label{conv_dia_cog}
\end{figure}

\begin{figure}[htb]
    \centering
      \begin{minipage}{0.48\columnwidth}
                       \footnotesize\centering\includegraphics[width=\columnwidth]{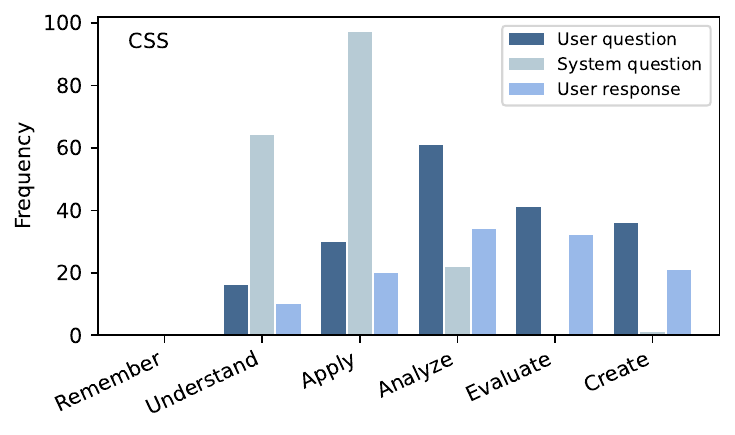}
    (a) Cognitive Level of User-LLM Dialogues in CSS
        \end{minipage}
        \begin{minipage}{0.48\columnwidth}
            \footnotesize\centering\includegraphics[width=\columnwidth]{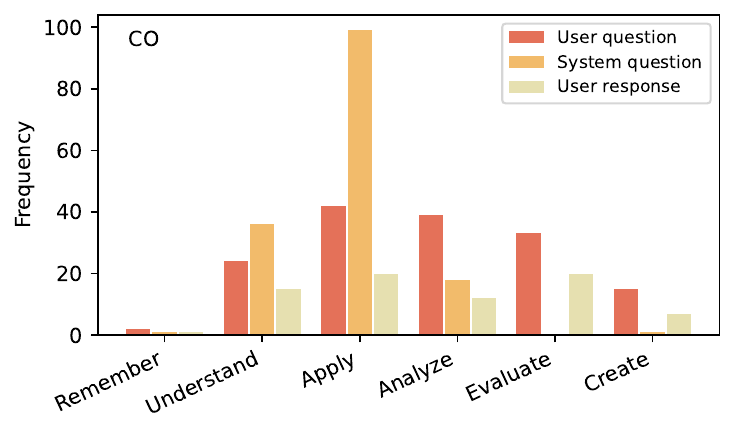}
    (b) Cognitive Level of User-LLM Dialogues in CO
        \end{minipage}
        
      \begin{minipage}{0.48\columnwidth}
                       \footnotesize\centering\includegraphics[width=\columnwidth]{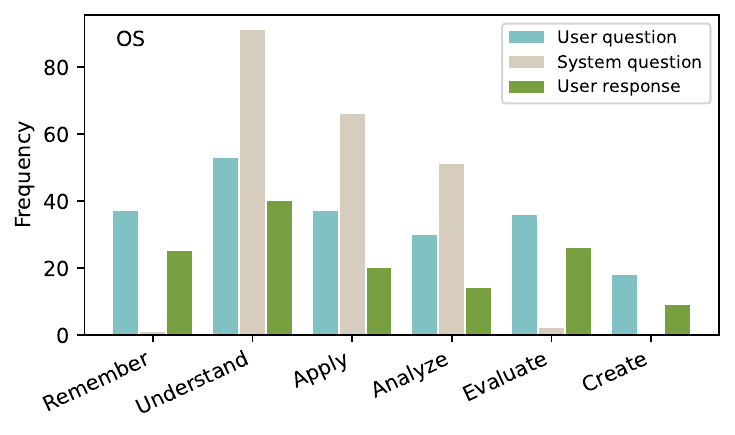}
    (c) Cognitive Level of User-LLM Dialogues in OS
        \end{minipage}

    \caption{Cognitive Level Classification of User-LLM Dialogues by Course} 
    \label{fig:conv_dia_cog_ana}
\end{figure}

A fundamental question for using LLMs as question initiators is whether they genuinely ask useful questions in an educational context. In general, for Computer Science for STEM and Computer Organization courses, the initial user questions are around the Analyze level, whereas those in Operating Systems average at the Apply level. The difference between the first user questions and the subsequent questions after being questioned by the LLMs is minor. Based on the classification results, there is no clear evidence that LLM-generated questions help deepen users' understanding or inspire higher-level educational exploration. Furthermore, LLM-generated questions are typically at the Apply level, which is either lower than or equivalent to the cognitive level of the users' questions. By nature, LLMs tend to ask questions related to explaining a knowledge concept or applying relevant knowledge to solve a problem. While these questions are useful, they are too focused on a particular level of cognitive ability. Immersion in such an interaction environment may negatively impact learners' cognitive development.


Considering our system design, these results imply that without further tuning, LLMs are unlikely to generate questions at higher cognitive levels. Human intervention remains crucial to fully facilitate the users' learning processes. It is worth discussing how to effectively prompt LLMs to generate diverse questions that meet all levels of cognitive ability requirements.

\section{Future Discussions}
We categorize the discussions from three primary perspectives: users, systems, and educators.

\subsection{Improvements towards current system}

At this stage, our system workflow is designed based on the assumption that user-system interactions adhere to an inquiry-based learning principle \cite{pedaste2015phases, abdi2014effect, friesen2013inquiry}. Inquiry-based learning posits that learners gain knowledge by formulating questions in response to challenging situations, deepening their understanding, and applying acquired knowledge to increasingly complex problems. Many specific features of our platform are grounded in this principle to enhance learning outcomes. For instance, the homework mode and its associated auto-detection feature serve to define and optimize the answer generation process for specific questions. Moreover, LLMs are prompted to generate follow-up questions at the conclusion of their responses to foster interactivity and encourage higher-order cognitive thinking. Although some features may not function as intended, the workflow remains well-structured and coherent.

Our data analysis identifies two primary obstacles that hinder the smooth progression of this workflow. First, the user-LLM interaction pattern diverges largely from user-human TA interactions. Users are expected to take initiative and self-regulate their learning process. Unlike traditional educational settings, where instructors can monitor and guide learning, self-regulation imposes substantial responsibility on users, who are typically students. Consequently, users with lower levels of self-motivation may derive less benefit from an LLM-powered educational system. Second, LLMs inherently possess distinct capabilities compared to human educators. Members of the teaching team can seamlessly access multimodal information when interacting with learners whereas LLMs face limitations in comprehensively obtaining and understanding diverse input formats. Additionally, LLMs tend to provide instructional explanations in a Socratic dialogue style. While LLMs are adept at presenting information due to their extensive training, members of the teaching team can adapt their teaching strategies more proactively and incorporate visual illustrations or practical examples as needed. Moreover, there is no clear evidence that LLMs possess the necessary abilities, such as mathematics or logical reasoning, required for a course assistant role \cite{mirzadeh2024gsm, parmar2024logicbench}.

To a certain extent, LLMs demonstrate a form of human-like intelligence. It's reasonable to design LLMs to emulate human behaviors, such as those of TAs, in an educational context. However, given the inherent differences between LLMs and human educators, users are interacting with fundamentally distinct teaching agents without fully recognizing these distinctions.

To address these challenges, several approaches can be considered. One approach is to inform users about the unique requirements of interacting with LLMs. However, such methods may inadvertently lead users to prefer general LLMs, such as ChatGPT, for convenience. A more pedagogically sound approach is to design an interaction system that serves as a bridge between users and LLMs. For example, we utilize the RAG technique to provide relevant course content as contextual background for LLM responses. While this process is straightforward for members of the teaching team who can effortlessly access multimodal information, it enhances the performance of LLM-powered course assistants, making them more closely resemble the members of the teaching team.

Nevertheless, these human-mimicry features come with inherent limitations. The RAG technique retrieves relevant data chunks based on similarity search, which may generally be effective but can prove inadequate in specific cases \cite{agrawal2024mindful, es2023ragas}. Furthermore, LLM hallucinations introduce confusion, as students may struggle to discern inaccuracies in the responses \cite{xu2024hallucination}. Additionally, LLM-generated questions often remain at a superficial cognitive level, raising concerns about their depth and quality. These limitations may hinder the effective integration of established pedagogical theories into LLM-powered systems.

All the limitations lead to an important question we are exploring regarding the integration of LLMs in education: What educational theories can be effectively implemented within LLM-powered educational platforms? Observing users engage in inquiry-based dialogues and receive detailed explanations is promising; it saves time and effort and potentially leads to better learning outcomes. However, such behavior resembles the functionality of advanced search engines more than that of proactive course assistants.

At present, inquiry-based learning is a well-established pedagogical principle that can support educational interactions between users and our system to some extent. Our preliminary findings indicate several challenges in successfully implementing this principle to its full extent. By nature, LLMs struggle to recognize the importance of applying educational theories in their responses. It remains an open question whether educational theories developed for human educators can be adapted for LLMs. Humans and LLMs represent fundamentally different types of teaching agents, each with overlapping but distinct inputs and outputs. Developing educational theories specifically tailored to human-LLM interactions represents a compelling area for future research.

\subsection{User: Agency, Initiative, and Learning Outcomes}

The sense of user agency provides users with the feelings and abilities necessary to control their learning environments and paths. Enhancing the sense of agency is a prevalent strategy in system design, as it can promote user engagement and encourage personalized learning experiences. Initiative describes a user's motivation to proactively engage with the system, fostering in-depth exploration of knowledge and higher-level cognitive learning. Providing excessive user freedom without sufficient pedagogical planning can lead to users feeling overwhelmed, which potentially stifles their initiative. Conversely, too little agency may make users feel restricted and diminish both their initiative and their overall learning outcomes. Striking a balance between these elements requires thoughtful system design. The key is to achieve a static or dynamic balance in which both structured planning and user autonomy are considered to guide users through a learning process. In this way, users are provided with sufficient flexibility to explore, decide, and take ownership of their learning processes.

Our results have revealed that approximately 21\% of user sessions result in no interactions; users open a chat but do not end up asking any questions. This may suggest that user initiative is not sufficiently supported or encouraged. Additionally, when users are asked questions by the LLMs, 65.27\% of these questions are not followed up by the users. We do not impose any restrictions to compel users to answer these questions, but it appears that users, particularly in higher-level courses, tend to ignore them. Such behavior significantly reduces the educational benefits that could be obtained by the users.

User initiative may also be influenced by the perceived learning outcome gained during the learning process. Even in basic inquiry-based learning settings, learners need to recognize the knowledge they have acquired in order to dive deeper into more advanced topics. The low response rate from users to LLM questions negatively impacts this learning loop. The lack of perceived learning achievement may further inhibit the effective progression of human-LLM interactions \cite{eom2006determinants}.

This naturally raises the question of how to strategically improve user agency and encourage students' learning initiative to optimize learning outcomes. Allowing students to select multiple modes for different educational scenarios, including providing hints rather than direct answers, offers them a higher level of agency when engaging with LLM-based course assistants. It is worth considering whether applying such strategies and granting LLM-based course assistants greater initiative to ask follow-up questions could improve learning outcomes. At the very least, keeping LLMs in a purely passive role misses the opportunity to fully realize their potential.

During the entire pilot study, we have observed that users were reluctant to share their dialogues using the provided share button on the web page or they instinctively neglected the existence of such an option. Rather than fostering greater connection and information exchange, users appear to be increasingly isolated from their peers when interacting with LLMs. This raises a concern that users' culture and social behaviors might gradually be affected. Some user questions contain basic grammatical errors and demonstrate a low level of question statement construction, which may be due to a lack of consideration for the human-like attributes of LLMs. The degradation of linguistic correctness in student questions raises concerns that LLM course assistants may provide biased responses \cite{harvey2025framework}. This behavior could potentially hinder student social growth.

\subsection{Educators: Culture and Engagements}
In LLM-powered educational systems, LLMs are increasingly assuming responsibilities and authority traditionally held by educators. Specifically, educators are gradually being excluded from the learning loop as LLMs become more involved in instructional processes. Our system has implemented several features intended to integrate educators into this paradigm by including an educator-curated knowledge database for RAG techniques, predefined course-level rules to regulate response generation, and an analytics page that provides insights into system usage and dialogue histories.

Despite these features, the system cannot fully leverage educators' pedagogical expertise. At present, educators primarily act as supervisors, monitoring system operations and ensuring users' psychological safety. Their extensive teaching expertise, accumulated through years of professional experience, remains largely untapped. This expertise could play a crucial role in guiding the LLM-powered systems to implement appropriate pedagogical theories and facilitate more meaningful and contextually rich responses.

Currently, our system has not successfully incorporated features that allow for such substantive educator involvement. The roles of educators remain restricted to basic administrative tasks, such as managing course documents. Enhancing educators' engagement is vital for further improving the usability and pedagogical effectiveness of our system \cite{harvey2024towards}. For instance, educators could contribute more course-related materials to enhance the scientific and educational context or provide sets of common questions with associated conditions to foster more interactive dialogues between LLMs and users. Enhanced educator engagement could lead to a more dynamic and adaptive learning system, fostering a healthier culture and a more effective learning environment \cite{harvey2025don,kizilcec2024advance}.

Moreover, deep educator engagement is foundational for achieving adaptive and even personalized learning experiences. Educational resources, developmental contexts, and appropriate pedagogical approaches vary significantly across regions and cultures. It makes a universal interface insufficient for integrating LLMs into education, and potentially exacerbates inequities. To address these challenges, we need to develop and implement a system design that actively encourages educators' participation and commitment to ensure that their expertise is effectively harnessed within the LLM-powered educational framework.

\section{Conclusion}

In this study, we develop and apply an LLM-powered educational system designed to function as a course assistant within educational contexts. The system is composed of multiple backend components that work in concert to support an open-ended, high-agency web interface. Users engage with this interface via chat-style interactions. By Spring 2024, the system has been deployed for approximately 2,000 students across six courses at three higher education institutions. We analyzed conversation data from three computer science courses in Spring 2024 to examine interaction patterns and evaluate educational value. Our findings are summarized as follows:

\textit{(A) Necessity and Effectiveness}: System usage metrics and the temporal distribution of interactions indicate a critical need for our system to address gaps left by traditional on-campus education. Overall, our system demonstrates a relatively high level of accuracy in supporting students' learning processes, thereby enhancing their educational experiences.

\textit{(B) Human-LLM Interaction Patterns}: Users typically take the initiative in their interactions with the system. Those in lower-level courses tend to seek more advanced cognitive activities from the LLMs and show a greater likelihood of responding to questions posed by the system.

\textit{(C) Challenges in Implementing Educational Theories}: Several system components are designed to align with the procedures of inquiry-based learning. However, our findings indicate that users exhibit limited interest in following these procedures. This raises concerns about the effective adaptation of traditional pedagogical theories to support new forms of human-LLM interaction. Furthermore, the current cognitive capabilities of LLMs are constrained when it comes to generating thought-provoking questions for users. While recent advancements in LLM research have led to significant achievements across various domains, the application of LLMs in educational contexts remains fraught with challenges. Researchers have developed numerous theoretical principles to facilitate effective learning, yet the application of these principles to human-LLM interactions requires further adaptation and evolution.

We discuss our findings from systematic, user, and educator perspectives. Currently, human-LLM interactions in education are predominantly centered around student-LLM dynamics, often to the detriment of incorporating educators' expertise. This imbalance poses significant obstacles to the development of advanced, realistic LLM-powered teaching systems that effectively serve the diverse needs of educational stakeholders.

\section*{Acknowledgments}
This material is based upon work partially supported by the U.S. National Science Foundation under Grant No. 2439815.


\section*{Declaration of competing interest}
Authors are current or former employees of HiTA AI Inc., a for-profit company.

\section*{Declaration of generative AI and AI-assisted technologies in the writing process}
During the preparation of this work the authors used ChatGPT \cite{openai2024chatgpt} and Claude \cite{anthropic2024claude} in order to improve language and readability. After using this tool/service, the authors reviewed and edited the content as needed and take full responsibility for the content of the publication.

\bibliographystyle{unsrt}  

\end{document}